\documentstyle[epsfig]{mn}

\def\spose#1{\hbox to 0pt{#1\hss}}
\def\lta{\mathrel{\spose{\lower 3pt\hbox{$\mathchar"218$}}
     \raise 2.0pt\hbox{$\mathchar"13C$}}}
\def\gta{\mathrel{\spose{\lower 3pt\hbox{$\mathchar"218$}}
     \raise 2.0pt\hbox{$\mathchar"13E$}}}
\def\<{{\langle}}
\def\>{{\rangle}}

\def\del{{\partial}}

\def\bO{{\bf \Omega}}

\def\eps{{\epsilon}}

\def\pc{{\rm \,pc}}
\def\kms{{\rm \,km \, sec^{-1}}}

\def\yr{{\rm \,yr}}
\def\NA{{\rm \,NA}}
\def\order{\mathcal{O}}

\newcommand{\grad}{\mbox{\boldmath $\nabla$}}
\newcommand{\vxi}{\mbox{\boldmath $\xi$}}
\newcommand{\hxi}{\mbox{\boldmath $\hat\xi$}}

\newcommand{\vx}{\mbox{\boldmath $x$}}
\newcommand{\bx}{\mbox{\boldmath $\bar x$}}

\newcommand{\vv}{\mbox{\boldmath $v$}}
\newcommand{\diver}{\mbox{\boldmath $\nabla\cdot$}}

\newcommand{\Lag}{{\cal L}}
\newcommand{\pd}{\partial}
\newcommand{\half}{\mbox{\small $\frac{1}{2}$}}
\newcommand{\BV}{Brunt-V\"ais\"al\"a~}
\newcommand{\He}{{\rm He}}
\newcommand{\boldk}{\mbox{\boldmath $k$}}
\newcommand{\boldv}{\mbox{\boldmath $v$}}
\newcommand{\boldx}{\mbox{\boldmath $x$}}
\newcommand{\boldhatx}{\mbox{\boldmath $\hat x$}}
\newcommand{\boldhaty}{\mbox{\boldmath $\hat y$}}
\newcommand{\boldhatz}{\mbox{\boldmath $\hat z$}}

\begin{document}

\title[Parametric instability]{
Linear and non-linear theory of a parametric instability of hydrodynamic
warps in Keplerian discs
}
\author[Charles F. Gammie, Jeremy Goodman and Gordon I. Ogilvie]
{Charles  F. Gammie,$^{1,2,3}$ Jeremy Goodman$^{1,4}$ and
Gordon I. Ogilvie$^{1,5,6}$\\
$^1$ Isaac Newton Institute, 20 Clarkson Road,
	Cambridge CB3 0EH \\
$^2$ Center for Astrophysics, MS-51, 60 Garden St., 
	Cambridge, MA 02138, USA  \\
$^3$ Center for Theoretical Astrophysics, University of Illinois, 
	1002 W. Green St., Urbana, IL 61801, USA \\
$^4$ Peyton Hall, Princeton, NJ 08544, USA \\
$^5$ Institute of Astronomy, Madingley Road, Cambridge CB3 0HA \\
$^6$ Max-Planck-Institut f\"ur Astrophysik,
  Karl-Schwarzschild-Stra\ss e 1,
  D-85740 Garching bei M\"unchen, Germany
}

\maketitle

\begin{abstract}
  We consider the stability of warping modes in Keplerian discs.  We
  find them to be parametrically unstable using two lines of attack,
  one based on three-mode couplings and the other on Floquet theory.
  We confirm the existence of the instability, and investigate its
  non-linear development in three dimensions, via numerical
  experiment.  The most rapidly growing non-axisymmetric disturbances
  are the most nearly axisymmetric (low $m$) ones.  Finally, we offer
  a simple, somewhat speculative model for the interaction of the
  parametric instability with the warp.  We apply this model to the
  masing disc in NGC 4258 and show that, provided the warp is not
  forced too strongly, parametric instability can fix the amplitude of
  the warp.
\end{abstract}


\section{Introduction}

Thin discs are an important model for accretion flows in a number of
different astrophysical objects, including active galactic nuclei,
interacting binary systems, and young stars.  Most disc models are
{\it flat}, i.e. the disc is planar, for simplicity and because there
has been little theoretical or observational motivation to consider
non-coplanar, or {\it warped} discs -- although observations of the
accreting neutron star system Her X-1 have long been interpreted as
requiring a precessing disc that lies out of the plane of the binary
system (Tananbaum et~al. 1972; Katz 1973; Roberts 1974).

A pair of developments in the last few years now strongly motivates
theoretical interest in warped discs.  First, Pringle (1996) found an
instability of a thin disc model that tends to produce large scale
warps.  The instability is driven by radiation torques exerted on the
disc as it is differentially illuminated by the central source and
then re-radiates.  Second, the megamaser in the nucleus of NGC 4258
has been resolved by the VLBA and can be understood as an almost
precisely Keplerian, but warped, thin disc.  The maser spots trace out
the position of the disc and allow accurate measurements of its
distance, the mass of the central object, and the parameters of the
warp (Miyoshi et~al. 1995).  Of course, warps may also be excited by
direct forcing, e.g. by tidal interaction of a T Tauri disc with a
binary companion (Papaloizou \& Terquem 1995).

Warps in unmagnetized, non-self-gravitating, inviscid Keplerian discs
are rather special, as we shall review in \S 3 below.  This is because
a warp (with azimuthal wavenumber $m=1$) that is quasi-static in an
inertial frame sets up vertically-varying horizontal pressure
gradients in the disc that oscillate at the rotation frequency
$\Omega$ as viewed in a locally corotating frame.  In Keplerian discs
this is resonant with the epicyclic frequency, and so modest warps can
excite large amplitude motions -- we shall call them {\it in-plane
  oscillations} -- in the disc.  As a result, warps and bending waves
propagate approximately non-dispersively, at nearly the sound speed,
in a Keplerian disc.  In a non-Keplerian disc, the in-plane
oscillations are much weaker and bending waves are dispersive
(Papaloizou \& Lin 1995; Ogilvie 1999).

The special nature of warps in Keplerian discs raises a number of
interesting astrophysical issues.  Are the in-plane oscillations
stable?  We shall show via two different lines of analysis that they
are not in \S 3 and \S 4.  For small warp amplitudes, the warp is
unstable to a parametric instability, a possibility presciently hinted
at by Papaloizou \& Terquem (1995).  Next, what is the non-linear
outcome of the instability?  This we explore numerically in \S 5,
which examines the immediate decay of an initially unstable warp.
Finally we discuss the implications of this for the disc of NGC 4258
in \S 6, using a simple model for the interaction of the warp with the
parametric instability.

\section{Basic model}\label{BASMOD}

To analyse this putative instability of warps in discs, we need to
formulate an equilibrium to perturb about.  In doing so we shall make
two seemingly severe approximations which we believe do not affect the
existence or nature of the instability.

First, we shall consider a radially local model of the disc, the well
known shearing-sheet model (Goldreich \& Lynden-Bell 1965).  In this
approach the equations of motion are expanded to lowest order in
$H(r)/r$ ($H$ being the disc scale height) about some point corotating
with the disc at $(r,\phi,z) = (r_0, \phi_0 + \Omega t, 0)$.  One then
erects a local Cartesian coordinate system
$(x,y,z)=\left(r-r_0,r_0(\phi-\phi_0-\Omega t),z\right)$ rotating with
uniform angular velocity $\bO=\Omega\boldhatz$.  In this limit, the
momentum equation is
\begin{equation}\label{LOCMODEOM}
{{\rm D}\boldv\over{{\rm D} t}} = -{1\over{\rho}}\grad P
 - 2 \bO \times \boldv + 2 q
\Omega^2 x \boldhatx - \Omega^2 z \boldhatz.
\end{equation}
The equilibrium velocity field is then $v_{y,0} = -q \Omega x$, where
$q \equiv - {\rm d}\ln\Omega/{\rm d}\ln r = 3/2$ for a Keplerian disc.  The
continuity equation takes its usual form, and we shall, unless
otherwise stated, assume an isothermal equation of state, $P =
c_{\rm s}^2\rho$.  The equilibrium density state is then $\rho = \rho_0
\exp(- z^2 / 2 H^2)$, with $H=c_{\rm s}/\Omega$.

Notice that we have neglected magnetic fields completely.  In a
magnetized disc turbulence will develop due to the Balbus-Hawley
instability (Balbus \& Hawley 1998), and the analysis presented here
will be invalid.  In that case one must study the interaction of the
resulting MHD turbulence with the warp, a project which, at present,
can only be undertaken numerically (Torkelsson et~al. 2000).  In discs
that are weakly magnetized the linear hydrodynamic analysis that
follows will still be applicable over a limited range of wavevectors,
although it seems likely that Balbus-Hawley initiated MHD turbulence
will ultimately dominate.  There may also be conditions under which
astrophysical discs are so poorly ionized that they do not couple to
the magnetic field (cf.  Gammie 1996; Gammie \& Menou 1998).

Our second approximation, which applies only to the analytical
approaches used below and not to the numerical experiments, is
axisymmetry for the bending waves and for the parasitic instabilities
that feed on them.  This may seem peculiar as the most astrophysically
interesting warps have $m = 1$; such modes have the unique property
that they can be stationary in an inertial frame, and furthermore are
what is observed in the unique megamaser disc in NGC 4258.
Nevertheless $m = 1$ modes are ``almost axisymmetric'' in that
$(1/r)\del_\phi \sim 1/r$, so it is consistent with the local
approximation to neglect terms of this order and consider only
axisymmetric bending waves for the basic state.  To the extent that
such terms are negligible, an $m=1$ mode with frequency $\omega=0$ and
an axisymmetric mode with $\omega=\Omega$ are locally
indistinguishable because both have the same frequency in a frame
corotating with the fluid.  That we consider only axisymmetric
parasitic instabilities on this basic state is perhaps a more
questionable approximation, but one that we relax in the non-linear
regime.

\section{Linear theory of Keplerian warps}\label{LINTHEORY}

To set the stage for the analysis that follows, we recount the theory
of axisymmetric waves in isothermal discs (Lubow \& Pringle 1993; see
also Goodman 1993; Ogilvie \& Lubow 1999).  We assume from the outset
that both the vertical structure of the disc and its dynamical
response are isothermal, so that $\gamma=1$ and the \BV frequency
vanishes.

Consider small, axisymmetric Eulerian perturbations $(\delta\vv,\delta
P)$ from the basic state, with time and space dependence $\exp[{\rm i} (k x
- \omega t)]$ understood.  These perturbations obey
\begin{equation}
\left({{\omega^2-\kappa^2}\over{{\rm i}\omega}}\right)\delta v_x=
-{\rm i}k\left({{\delta P}\over{\rho}}\right),
\end{equation}
\begin{equation}
-{\rm i}\omega\,\delta v_z=-\del_z\left({{\delta P}\over{\rho}}\right),
\end{equation}
\begin{equation}
-{\rm i}\omega\left({\delta P\over{\rho}}\right)=\Omega^2 z\,\delta
v_z-c_{\rm s}^2({\rm i}k\,\delta v_x+\del_z\delta v_z).
\end{equation}
Here $\kappa$ is the epicyclic frequency, given by
$\kappa^2=(4-2q)\Omega^2$.  One can then obtain a single second-order
equation for the pressure perturbation,
\begin{eqnarray}
\lefteqn{\del_z^2\left({{\delta P}\over{\rho}}\right)-
\left(\frac{z}{H^2}\right)\del_z\left({{\delta P}\over{\rho}}\right)}&
\nonumber\\
&&+\frac{\omega^2}{c_{\rm s}^2}\left[1-\left(\frac{c_{\rm s}^2k^2}
{\omega^2-\kappa^2}\right)\right]\left({{\delta P}\over{\rho}}\right)=0.
\end{eqnarray}
The eigenfunctions are Hermite polynomials\footnote{For $\gamma\ne 1$,
  the solutions can also be expressed in terms of Hermite polynomials,
  but with arguments depending on $k$ and $\omega$.} (e.g. Abramowitz
\& Stegun 1965),
\begin{equation}
{{\delta P}\over{\rho}}\propto\He_n(\zeta)\equiv
(-1)^n {\rm e}^{\zeta^2/2}\frac{{\rm d}^n}{{\rm d}\zeta^n}
{\rm e}^{-\zeta^2/2},
\end{equation}
where $\zeta=z/H$ and $n$ is a non-negative integer.  We recall their
differential equation
\begin{equation}
\He_n''(\zeta)-\zeta\He_n'(\zeta)+n\He_n(\zeta)=0
\end{equation}
and orthogonality relation
\begin{equation}
\int\limits_{-\infty}^{\infty}{\rm e}^{-\zeta^2/2}\He_m(\zeta)\He_n(\zeta)
\,{\rm d}\zeta=n!\sqrt{2\pi}\delta_{mn}.
\end{equation}
The corresponding eigenfrequency $\omega_n$ therefore satisfies
\begin{equation}\label{dispersion}
\frac{\omega_n^2}{\Omega^2}\left[1-\left(\frac{c_{\rm s}^2k^2}
{\omega_n^2-\kappa^2}\right)\right]=~n~\in\{0,1,2,3,\ldots\}.
\end{equation}
For future reference we also note the corresponding Lagrangian
displacement,
\begin{eqnarray}\label{solution}
\xi_{x,n}&=&{\rm i}kH\left(\frac{\omega_n^2}{\omega_n^2-\kappa^2}\right)
{\cal N}_n\He_n(\zeta),\nonumber\\
\xi_{y,n}&=&2kH\left(\frac{\omega_n\Omega}{\omega_n^2-\kappa^2}\right)
{\cal N}_n\He_n(\zeta),\nonumber\\
\xi_{z,n}&=& {\cal N}_n\He_n'(\zeta)={\cal N}_n n\He_{n-1}(\zeta),
\end{eqnarray}
where the normalization
\begin{equation}\label{normfactor}
{\cal N}_n=\left(\frac{2}{n!}\right)^{1/2}
\left[n+\left(\frac{kH\omega_n^2}
{\omega_n^2-\kappa^2}\right)^2\right]^{-1/2}\,H
\end{equation}
is chosen according to
\begin{equation}\label{massnorm}
\int \rho(\xi_x^2 + \xi_z^2)\, {\rm d}^3\boldx = M H^2,
\end{equation}
where
\begin{equation}
M=\int\rho\, {\rm d}^3\boldx
\end{equation}
is the mass of the fluid.  The eigenfunction $\vxi_n$ has $n$ vertical
nodes in $\xi_x$ and $n-1$ nodes in $\xi_z$.  The dispersion relation
(\ref{dispersion}) has two solutions (both positive) for $\omega_n^2$
at each $n$: an ``acoustic'' branch with $\omega_n^2\ge
c_{\rm s}^2k^2+\kappa^2$ and an ``inertial'' branch with $\omega_n^2
\le\kappa^2$.  The bending waves correspond to $n=1$, and the density
wave, for which $\xi_z=0$ and $\del_z\xi_x=\del_z\xi_y=0$, lies on the
acoustic branch at $n=0$.

In the special case of a Keplerian disc, $\kappa=\Omega$, the
bending-wave branches cross over at $k=0$ with non-vanishing group
velocities $\pm c_{\rm s}/2$.  This happens because of the coincidence of
the natural frequencies of horizontal and vertical oscillations.  A
disturbance with long wavelength, $kH \ll 1$, is simultaneously close
to the Lindblad and vertical resonance conditions.  The form of the
dispersion relation for small $kH$ is
\begin{equation}
\omega\approx\Omega\pm\frac{1}{2}c_{\rm s}k,
\end{equation}
and the normalized eigenfunction in this limit is
\begin{equation}
\vxi=(\pm {\rm i}z,\pm2z,H).
\end{equation}
The physical velocity field, with arbitrary normalization, is
\begin{equation}
\delta\vv=S\left(z\cos(\Omega t),-\frac{1}{2}z\sin(\Omega t),\mp
H\sin(\Omega t)\right),
\label{deltav}
\end{equation}
where 
\begin{equation}
S \equiv \left| {{\rm d} \delta v_x\over{{\rm d}z}} \right|
\end{equation}
is a measure of the shear rate in the warp.

This velocity perturbation consists of two parts.  The vertical motion
is simply a uniform oscillation at the vertical frequency (which, in a
Keplerian disc, is equal to the orbital frequency).  In a disc with an
$m=1$ warp,\footnote{We recall that the same eigenfunctions apply to
  non-axisymmetric modes of small $m$ ($m\ll r/H$) provided that
  $\omega$ is understood as the frequency seen in a locally corotating
  frame.} it corresponds to the fluid following an inclined circular
orbit.  This motion can be eliminated by redefining the frame of
reference to follow the inclined orbit, with the result that only the
horizontal components of (\ref{deltav}) remain.  The horizontal motion
is in fact an {\it exact} solution to the equations of motion in the
shearing sheet, so $S$ can take on any value.  We shall consider the
stability of this motion; it is in a sense a local model in azimuth
and radius for the motion in a large scale, $m = 1$, warp.  For the
special case of a quasi-static $m=1$ warp in a Keplerian disc, the
disturbance can remain close to the Lindblad-vertical resonance over
the entire radial extent of the disc.

In the limit that the amplitude of the warp is very large, so that $S
\gg \Omega$, the oscillatory character of the motion is not important
and the problem reduces to the stability of a vertically linear shear
flow.  This problem was considered by Kumar \& Coleman (1993), who
found it unstable.

\section{Stability of warps: Floquet analysis}\label{FLOQUET}

The stability problem is reduced to its simplest possible form if we
assume that the fluid is incompressible, rather than isothermal, and
neglect its stratification.  The governing equations are the same as
before, except that $\grad\!\cdot\!\boldv = 0$ and the vertical
gravity term is omitted in the equation of motion.

The basic state is now regarded as infinite in both radial and
vertical extent, and has uniform density and pressure.  We divide the
velocity field into three pieces: $\vv = \vv_0 + \vv_1 + \delta\vv$.
The first piece, $\vv_0= -{3\over{2}}\Omega x \boldhaty$, is the
differential rotation of the disc as it appears in the local model;
the second piece is the ``bending wave'', given by the following exact
solution to the equations of motion (cf. equation \ref{deltav}):

\begin{equation}\label{vxbend}
v_{x,1} = S z \cos(\Omega t),
\end{equation}
\begin{equation}
v_{y,1} = -{1\over{2}} S z \sin(\Omega t),
\end{equation}
and $v_{z,1} = 0$.  Since the solution is exact, $S$ can be
arbitrarily large.

The third piece of the velocity field, $\delta\vv$, is the
perturbation on the warp whose stability we wish to consider.  In
general, one might think of linearizing the equation of motion and
projecting $\delta\vv$ on to a suitable basis of functions.  Note that
one cannot find normal modes because the basic flow is itself
time-dependent.  However, a standard technique is available for an
incompressible fluid if the strain tensor of the basic flow is
independent of position (Kelvin 1887; Craik \& Criminale 1986; Bayly
1986).  Here one considers a perturbation in the form of a Fourier
mode whose wavevector evolves in time according to the strain field.
Specifically, we take all perturbed quantities to be time-dependent
amplitudes multiplied by $\exp[{\rm i} \boldk(t) \cdot \boldx]$, where
\begin{equation}
{{\rm d} \boldk\over{{\rm d} t}} = - [\nabla(\vv_0+\vv_1)]\cdot\boldk.
\end{equation}
For an axisymmetric mode ($k_y=0$) this gives
\begin{equation}
k_x={\rm constant},\qquad{{\rm d}k_z\over{{\rm d} t}} = - S k_x \cos(\Omega t),
\end{equation}
so that $k_z = k_{z,0} - (S/\Omega) k_x \sin(\Omega t)$.  In words,
then, these perturbing waves nod back and forth on top of the
oscillating in-plane motions set up by the warp.  In the absence of
the warp, they would be inertial oscillations (also called r modes)
with $\omega^2 = \Omega^2 k_z^2/k^2$.

The amplitudes of the perturbations obey
\begin{equation}
{{\rm d} \delta v_x\over{{\rm d} t}} - 2\Omega \,\delta v_y +
S\cos(\Omega t) \delta v_z = -{\rm i} k_x \left({{\delta P}\over{\rho}}\right),
\end{equation}
\begin{equation}
{{\rm d} \delta v_y\over{{\rm d} t}} + {1\over{2}}\Omega \,\delta v_x -
{1\over{2}}S\sin(\Omega t) \delta v_z = 0,
\end{equation}
\begin{equation}
{{\rm d} \delta v_z\over{{\rm d} t}}  =
-{\rm i} k_z \left({{\delta P}\over{\rho}}\right),
\end{equation}
\begin{equation}
{\rm i}k_x\,\delta v_x+{\rm i}k_z\,\delta v_z = 0.
\end{equation}
Note that this is an exact solution since the non-linear
self-interaction of the mode is zero.

Recall that $k_z$ is time-dependent.  These equations can be reduced
to a single Floquet equation (a linear equation with periodic
coefficients) of the form
\begin{equation}
{{\rm d}^2\over{{\rm d} t^2}}\left( k^2 k_z^{-1/2} \delta v_z \right) + 
	f(t)\left( k^2 k_z^{-1/2} \delta v_z \right) = 0,
\label{floquet}
\end{equation}
where
\begin{equation}
f(t) \equiv {k_z^2 \Omega^2\over{k^2}} - {3 \dot{k_z}^2\over{4 k_z^2}}
	+ {\ddot{k_z}\over{2 k_z}} +
	{k_z \ddot{k_z}\over{k^2}}
\end{equation}
is a function with period $2\pi/\Omega$.  

First we assume that the shearing motion is small, in that $S/\Omega
\ll 1$.  Expanding $f$ in powers of $S/\Omega$,
\begin{equation}
f(t) = {k_{z,0}^2\over{k_0^2}}\Omega^2 + {k_x (k_x^4 + 3
k_{z,0}^4)\over{2 k_{z,0} k_0^4}} \Omega S \sin(\Omega t) +
O\left({S\over{\Omega}}\right)^2.
\end{equation}
To this order the equation for the evolution of the perturbation
is the Mathieu equation, which has the normal form 
\begin{equation}
{{\rm d}^2 y\over{{\rm d} t^2}} + \Omega^2 (a + 2 \eps \cos \Omega t) y = 0
\end{equation}
(see, e.g., Bender \& Orszag 1978).  

The Mathieu equation is the classical model equation for {\it
  parametric instability}.  If $\eps \ll 1$, parametric instability
sets in for $a = n^2/4$, $n = 0, 1, 2, \ldots$.  Here the Mathieu
equation describes the evolution of an inertial oscillation with
initial wavevector $\boldk_0$ and frequency $< \Omega$, so $a < 1$.
Instability is then achieved by tuning $\boldk_0$ so that $n = 1$, or
equivalently, so that the inertial oscillation has frequency $\pm
\Omega/2$.  The inertial oscillation is destabilized by its
interaction with the warp, which has frequency $\Omega$.  Near $n = 1$
the growth rate of the instability is $\eps\Omega$.

Applying these results to the warp, we find instability when
$k_x/k_{z,0} = \pm \sqrt{3}$, so the wavevector of the perturbed
velocities is initially oriented at $\pm 60\degr$ to the
vertical.  Then
\begin{equation}
f(t) = \Omega^2 \left({1\over{4}} \pm {3\sqrt{3}\over{8}}{S\over{\Omega}}
	\sin(\Omega t)\right) + O\left({S\over{\Omega}}\right)^2.
\end{equation}
The growing mode has growth rate
\begin{equation}
s = {3\sqrt{3}\over{16}} S + O\left({S\over{\Omega}}\right)^2,
\label{small_s}
\end{equation}
i.e. the growth rate is proportional to the shear rate in the warp.

If the shearing motion is not small, the Floquet exponents and
associated growth rate can still be obtained numerically.  Instead of
using equation (\ref{floquet}), which is singular if $S$ is so large
that $k_z$ passes through zero, one can integrate the coupled
equations
\begin{eqnarray}
{{{\rm d}\delta v_z}\over{{\rm d}t}}&=&k_z\left(-{{{\rm i}\,\delta
      P}\over{\rho}}\right),\nonumber\\
{{{\rm d}}\over{{\rm d}t}}\left(-{{{\rm i}\,\delta
      P}\over{\rho}}\right)&=&-\left({{k_z\Omega^2+3\ddot
      k_z}\over{k^2}}\right)\delta v_z\nonumber\\
    &&\qquad-{{4k_z\dot k_z}\over{k^2}}
      \left(-{{{\rm i}\,\delta P}\over{\rho}}\right)
\end{eqnarray}
over a single period, starting from two linearly independent initial
conditions.  The result of this calculation is that, as $S$ increases,
the growth rate $s$ increases less rapidly than a simple linear
relation.  For $S/\Omega=0.1$, $1$ and $10$, the growth rate
(maximized with respect to $k_x/k_{z,0}$) is $99.9\%$, $90.1\%$ and
$31.1\%$, respectively, of the value given in equation
(\ref{small_s}).

We remark that this calculation could easily be extended to
non-axisymmetric modes, and viscosity could also be allowed for.
However, the method is restricted to incompressible fluids.
 
In the case of a non-Keplerian disc ($\kappa\neq\Omega$) the analysis
is almost identical.  Provided that $S$ still measures the amplitude
of the $x$-component of the epicyclic motion, $v_{x,1}=Sz\cos(\kappa
t)$, the Floquet equation and growth rate are unchanged except that
$\Omega$ is replaced by $\kappa$.  The main difference in the case of a
non-Keplerian disc is that the in-plane oscillations are more weakly
coupled to the warp.

This simple calculation strongly suggests that warped discs are
hydrodynamically unstable, although it uses a model lacking
compressibility and the vertical structure of the disc.  In the next
section we remove these limitations, at the price of a lengthier and
less transparent analysis.

\section{Stability of warps: mode coupling analysis}

Parametric instability can be treated as a special case of three-mode
coupling in which one of the three modes (mode 1) has a prescribed
motion, and hence can be regarded as a time-dependent component of the
basic state upon which the other two modes (2 and 3) are treated as
linear perturbations.\footnote{The amplitude of mode 1 must still be
  small, $\order(\eps)$.  Our expansion (equations [37,38]) assumes
  that the amplitude of modes 2 and 3 is $\order(\delta\eps)$, with $1
  \gg \delta \gg \eps$.}  Sometimes, as is the case here, the
wavelength of mode 1 is much longer than that of mode 2 or 3.

In the absence of dissipation, three-mode couplings are most easily
calculated from an action principle,
\begin{equation}\label{action}
\delta_\xi\int {\rm d}t \int {\rm d}^3\bx
\,\Lag(\vxi,\pd_t\vxi,\pd_i\vxi,\bx,t)=0.
\end{equation}
Here $\delta_\xi$ indicates a first-order variation with respect to
the displacements $\vxi$, which are regarded as functions of the
spatial coordinates $\bx$ and time $t$.  Thus $\vx=\bx+\vxi(\bx,t)$ is
the actual position of a fluid element whose equilibrium position is
$\bx$.  For clarity, we sometimes indicate the basic state with an
overbar: e.g., $\bar\rho(\bx)$ is the mass density in the basic
state.  In the case of axisymmetric motions of the shearing sheet, the
Lagrangian density to be used in equation (\ref{action}) is
\begin{equation}\label{SSLag}
\Lag= \bar\rho(\bar z)\left\{\frac{1}{2}\left[(\pd_t\xi_x)^2 
+(\pd_t\xi_y)^2 -\kappa^2\xi_x^2-\Omega^2\xi_z^2\right]
~-U\right\}.
\end{equation}
The frequencies $\Omega$ and $\kappa$ are constants.  For simplicity,
we assume that $U$, the internal energy per unit mass, has the form
appropriate to an isothermal gas of uniform sound speed $c_{\rm s}$:
\begin{equation}\label{polytrope}
U= c_{\rm s}^2\ln\rho.
\end{equation}
The density $\rho$ differs from $\bar\rho$ because of changes in
volume:
\begin{eqnarray}\label{Jacobian}
\left(\frac{\rho}{\bar\rho}\right)^{-1}&=&
 \det\left(\delta_{ij}+\frac{\pd\xi^i}{\pd\bar x^j}\right)\nonumber\\
&=& 1 \,+\diver\vxi ~+(\pd_x\xi_x)(\pd_z\xi_z)
-(\pd_z\xi_x)(\pd_x\xi_z).
\end{eqnarray}
Therefore, the internal energy (\ref{polytrope}) contains terms of all
orders in $\vxi$.  The quadratic and cubic terms are
\begin{eqnarray}\label{TaylorU}
U_2 &=&c_{\rm s}^2\left[\frac{1}{2}(\diver\vxi)^2 + 
\pd_x\xi_z\pd_z\xi_x-\pd_x\xi_x\pd_z\xi_z\right],\label{U2}\\
U_3 &=& c_{\rm s}^2\left[(\diver\vxi)(\pd_x\xi_x\pd_z\xi_z-
\pd_x\xi_z\pd_z\xi_x)-\frac{1}{3}(\diver\vxi)^3\right].
\label{U3}
\end{eqnarray}
If the spatial derivatives of $\vxi$ are small, higher-order terms can
be neglected.  Then $U_2$ and the other quadratic parts of $\cal L$
determine the linear equations of motion, while $U_3$ provides the
dominant non-linear interactions.

A general disturbance can be expanded in terms of the spatial
eigenfunctions $\{\hxi_\alpha\}$ of the linear normal
modes:\footnote{This assumption would be highly questionable for
  non-axisymmetric perturbations: in that case, the modes are probably
  not complete, or if complete, not discrete.}
\begin{equation}\label{expand}
\vxi(\bx,t)= \sum\limits_\alpha q_\alpha(t)\hxi_\alpha(\bx).
\end{equation}
Absent non-linear interactions, the coefficients $\{q_\alpha(t)\}$
evolve independently according to
\begin{equation}\label{harmonic}
\ddot q_\alpha + \omega_\alpha^2 q_\alpha =0,
\end{equation}
where $\omega_\alpha$ is the natural frequency of the $\alpha^{\rm
  th}$ normal mode.  Non-linear equations of motion for the $q$s result
from
\begin{displaymath}
\delta_q\int {\rm d}t\, L\left(\{q_\alpha\},\{\dot q_\alpha\}\right)=0,
\end{displaymath}
with
\begin{eqnarray}\label{Lq}
L&\equiv &\int {\rm d}^3\bx\, \Lag(\vxi,\pd_t\vxi,\pd_i\vxi,\bx,t)\nonumber\\
&=& \sum\limits_\alpha\frac{MH^2}{2}\left(\dot q_\alpha^2-\omega_\alpha^2
q_\alpha^2\right)\nonumber\\
&&-\int {\rm d}^3\bx\,\bar\rho(\bx)\,
U_3\left(\sum\limits_{\alpha} q_\alpha(t)\hxi_\alpha(\bx)\right)
~+O(q^4).
\end{eqnarray}
The spatial eigenfunctions have the normalization (\ref{massnorm});
therefore $\hxi_\alpha$ has units of length and $q_\alpha$ is
dimensionless.

Consider a situation in which only three modes $\{q_1,q_2,q_3\}$ are
significantly excited.  Then the Lagrangian reduces to
\begin{equation}\label{threeL}
L = \sum\limits_{\alpha=1}^3 \frac{1}{2}(\dot q_\alpha^2-\omega_\alpha^2
 q_\alpha^2) ~-\Gamma q_1 q_2 q_3,
\end{equation}
where we have divided out an overall factor of $MH^2$, and
\begin{eqnarray}\label{coupling}
\lefteqn{
\Gamma = \frac{1}{MH^2} \left[\frac{\pd^3}{\pd q_1\pd q_2\pd q_3}~
\int {\rm d}^3\bx\,\bar\rho(\bar z)\,
U_3\left(\sum\limits_{\alpha} q_\alpha(t)\hxi_\alpha(\bx)\right)
\right]_{q=0}.
}&\nonumber\\
\end{eqnarray}

The behaviour of a system of three oscillators non-linearly coupled as
in (\ref{threeL}) is well known \cite{sg69,kg96}:

\noindent
1. The Hamiltonian corresponding to (\ref{threeL}) has a local but not
a global minimum at the origin, because the highest terms are odd in
$q$.  When, as here, (\ref{threeL}) results from truncation of a
physical system with a positive-definite Hamiltonian, then quartic and
perhaps higher terms must be retained if the amplitude or coupling is
sufficiently large.

\noindent
2. In the opposite limit of small amplitudes or weak coupling, where
the cubic term is small compared to the quadratic ones, non-linearity
is important only if approximate resonance exists among the natural
frequencies of the three linearized oscillators.  Without loss of
generality\footnote{We presume that the linearized oscillators are
  stable, so that every $\omega_\alpha^2\ge 0$, and we define
  $\omega_\alpha\equiv |\sqrt{\omega_\alpha^2}|$}, let
$\omega_1\ge\omega_2\ge\omega_3$.  Resonance exists if
$\omega_1\approx\omega_2+\omega_3$, and the accuracy of the resonance
is characterized by the smallness of
\begin{equation}\label{resonance}
\Delta\omega\equiv\omega_1 -\omega_2-\omega_3.
\end{equation}
On the other hand, the strength of the coupling can be characterized
by a dimensionless parameter such as
\begin{equation}\label{strength}
\epsilon\equiv \left|\frac{\Gamma E^{1/2}}{\omega_1^3}\right|,
\end{equation}
where $E$ is the total energy of the system (\ref{threeL}).

\noindent
3. For $\epsilon\ll 1$, slow exchange of energy among the three
oscillators can occur on time-scales $\sim(\epsilon\omega)^{-1}$
provided $|\Delta\omega|\lta|\epsilon\omega_1|$.  There is no general
restriction on the amount of energy that can be exchanged, except that
\begin{equation}\label{selection}
\frac{\Delta E_3}{\omega_3}=\frac{\Delta E_2}{\omega_2}
=-\frac{\Delta E_1}{\omega_1},
\end{equation}
and that the energy in each oscillator,
\begin{equation}\label{oscen}
E_\alpha\equiv \half\left(\dot q_\alpha^2+\omega_\alpha^2 q_\alpha^2
\right),
\end{equation}
must remain positive.  The selection rule (\ref{selection}) can be
explained, or at least remembered, by saying that a pair of quanta of
action $I=E/\omega$ in oscillators $2$ and $3$ combine in pairs to
make a single quantum in oscillator $1$.

\noindent
4. In the special case that all of the energy is initially in
oscillator 1, the energies of oscillators 2 and 3 grow exponentially
at growth rate $2s$, where $s$ is the growth rate of their amplitudes.
The maximum growth rate is achieved when $\Delta\omega=0$:
\begin{equation}\label{maxrate}
s_{\rm max}=|\Gamma|\sqrt{\frac{E_1}{8\omega_1^2\omega_2\omega_3}}
\quad=\frac{\dot E_2}{2E_2}=\frac{\dot E_3}{2E_3}.
\end{equation}
Of course growth slows once a significant fraction of the total energy
resides in oscillators 2 and 3.  If resonance is not exact, then the
growth rate is
\begin{equation}\label{genrate}
s =\sqrt{s_{\rm max}^2-(\Delta\omega/2)^2},
\end{equation}
and there is no growth at all where $|\Delta\omega|\ge 2 s_{\rm max}$.

\noindent
5. Finally, if dissipation is present so that the free decay rate of
the amplitudes of oscillators 2 and 3 is $\eta$, then formula (\ref{genrate})
holds with $s_{\max}$ replaced by $s_{\max}-\eta$.

Finding the parametric growth rate due to a large-scale bending mode
reduces therefore to the evaluation of the coupling parameter $\Gamma$
given by equation (\ref{coupling}) in terms of the linear eigenfunctions
and eigenfrequencies $\{(\hxi_\alpha,\omega_\alpha)\}$.  In a local
approximation, the background density $\bar\rho$ depends only upon
$\bar z$, so the $(\bar x,\bar z)$ dependences are separable.
Hereafter we omit the overbar from quantities of the basic state
except where needed for clarity.  Thus the horizontal dependence of
the eigenfunction is $\exp(ik_\alpha x)$.  To obtain a non-zero result
for the coupling (\ref{coupling}), we must have
\begin{equation}\label{ktriad}
k_1+k_2+k_3=0.
\end{equation}
In most cases of interest, mode 1 is a large-scale bending mode with
$k_1\ll H^{-1}$, and modes 2 and 3 have wavelengths $\lta 2\pi H$, so
to an adequate approximation, $k_2\approx -k_3\gg k_1$, and we shall
write $k$ for $k_2$.  In the limit $k_1=0$, assuming that
$\kappa=\Omega$, the bending mode is (cf. \S 3)
\begin{equation}\label{bend}
(\xi_x,\xi_y,\xi_z)_1 = \frac{S}{\Omega}
\mbox{Re}\left[(z,2{\rm i}z,{\rm i}H) {\rm e}^{-{\rm i}\Omega t}\right],
\end{equation}
where $S$ is the vertical shear of the mode as defined by
equation (\ref{vxbend}).

Comparing equations (\ref{bend}) and (\ref{U3}), one sees that the coupling
is dominated by terms involving $(\pd_z\hat\xi_x)_1$, i.e. by the
vertical strain due to the bending mode.  Retaining only the leading
order terms,
\begin{eqnarray}\label{cupapprox}
\lefteqn{
\Gamma\approx -\frac{1}{\Sigma H^2}\int {\rm d}z\,\bar\rho(z)c_{\rm s}^2
\left\langle(\diver\hxi)_2(\pd_x\hat\xi_z)_3 +
(\diver\hxi)_3(\pd_x\hat\xi_z)_2\right\rangle_x,
}\nonumber\\
\end{eqnarray}
where $\Sigma\equiv\int {\rm d}z\,\bar\rho(z)$ is the surface density.

Whereas the eigenfunction (\ref{bend}) of the bending wave is
independent of the vertical structure of the disc in the
long-wavelength limit, the eigenfunctions of the short-wavelength
daughter modes 2 and 3 are not.  These are described in \S
\ref{LINTHEORY}.

We are now in a position to calculate the coupling (\ref{cupapprox}).
Note that modes 2 and 3 must have different values of $n$: under the
reflection $z\to-z$, $\diver\hxi_n$ has the parity $(-)^n$, whereas
$\pd_x\hat\xi_z$ has the parity $(-)^{n-1}$. Therefore $\Gamma\ne0$
only if $n_2-n_3$ is odd.  This is probably true for any vertical
structure symmetric about the mid-plane, if $n$ counts vertical nodes.
For the isothermal gas, the properties of the Hermite polynomials are
such that $\Gamma\ne0$ requires $|n_3-n_2|=1$.  We must also remember
to change the sign of $k$ between the two modes, because $k\equiv
k_2\approx -k_3$.  Finally, it turns out that the horizontal phase of
modes 2\& 3 must differ by $90^\circ$ (a factor ${\rm i}$ in our complex
notation) in order to get the maximum coupling.  Taking
$\kappa=\Omega$, we find
\begin{eqnarray}\label{Gamres}
\Gamma&=&kH\sqrt n\omega_{n-1}^2
\left[n+\left(\frac{kH\omega_n^2}{\Omega^2-\omega_n^2}\right)^2\right]^{-1/2}
\nonumber\\
&&\qquad\times
\left[n-1+\left(\frac{kH\omega_{n-1}^2}{\Omega^2-\omega_{n-1}^2}\right)^2
\right]^{-1/2}.
\end{eqnarray}
Here $\omega_n$ and $\omega_{n-1}$ are implicit functions of $k$ and
$n$ through equation (\ref{dispersion}) and the resonance condition
$\omega_n+\omega_{n-1}\approx\Omega$.

For comparison with results obtained in \S \ref{FLOQUET}, we
consider the limit
\begin{eqnarray*}
kH&\to&\infty,\\
n&\to&\infty,\\
\omega_n&\to&\omega_{n-1}\to\frac{1}{2}\Omega.
\end{eqnarray*}
It is easy to see from (\ref{dispersion}) that this requires
$kH/\sqrt{3n}\to 1$, and then we obtain
\begin{equation}\label{Glimit}
\Gamma\to\Gamma_\infty= \frac{3\sqrt{3}}{16}\Omega^2.
\end{equation}

Taking into account that the total energy of the harmonic oscillation
(\ref{harmonic}) is twice its mean kinetic energy, we have
$E_1=\Omega^2 |q_1|_{\max}^2/2$, where $|q_1|_{\max}=|S/\Omega|$ is
the dimensionless semi-amplitude of the bending mode.  The growth rate
(\ref{maxrate}) reduces to
\begin{equation}\label{finals}
s_{\max}= \frac{|\Gamma S|}{4\Omega\sqrt{\omega_2\omega_3}}.
\end{equation}

In particular, for $\Gamma\to\Gamma_\infty$ and
$\omega_2=\omega_3=\half\Omega$, $s_{\max}=3\sqrt{3}S/32$.  This is
exactly half the growth rate calculated in \S \ref{FLOQUET}.  The
explanation appears to be that each daughter mode with $n\gg 1$ and
$k\approx \sqrt{3n}$ can participate in two parametric couplings: one
with mode $(n-1,-k)$ and the other with $(n+1,-k)$.  At sufficiently
large $n$, both of the resonance conditions $\omega_{n,k}+\omega_{n\pm
  1,-k}\approx\Omega$ can be satisfied with nearly equal accuracy.  As
a result, the growth rate of mode $(n,k)$ is doubled.

It is, of course, not obvious that the mode-coupling calculation
should give the same answer as the Floquet analysis of \S4, even after
this correction.  The gas considered here is compressible, whereas in
\S4 it was assumed incompressible.  As $n\to\infty$, however, the
modes of interest, which belong to the ``inertial'' branch (\S3), are
effectively incompressible near the mid-plane.  At high altitudes ($z
> H$) the inertial modes are nearly isobaric, and the local
contribution to the coupling coefficient (\ref{coupling}) varies with
height.  But when modes of many (large) values of $n$ and the same
$|k|$ grow together, it can be shown that the relative phasing of the
mode amplitudes demanded by parametric instability results in a total
wavefunction that is concentrated toward the mid-plane.

\section{Non-linear outcome}

The analyses of \S\S 4 and 5 establish the existence of a linear
instability of hydrodynamic warps.  We would now like to understand
the non-linear outcome of the instability, and its non-axisymmetric
development, which was ignored in the analytic analysis.  To do this,
we have integrated the hydrodynamic equations in a series of numerical
experiments.

\subsection{Initial conditions, boundary conditions, and numerical method}

The numerical model retains the basic assumptions used in the linear
analysis.  The parent warp mode is axisymmetric.  The equations of
motion are those appropriate to the local model (see equation 1).  The
disc is Keplerian, so $q = 3/2$.  The fluid is assumed strictly
isothermal.  We do not require that the system remain axisymmetric.

In the initial conditions we set
\begin{equation}
v_x = S z + \delta,
\end{equation}
\begin{equation}
v_y = -(3/2)\Omega x + \delta,
\end{equation}
\begin{equation}
v_z = 0 + \delta,
\end{equation}
\begin{equation}
\rho = \rho_0 \exp(-z^2/(2 H^2)).  
\end{equation}

Here $|\delta| < 10^{-3}$ is a uniformly distributed random variable
with mean zero chosen independently for each variable in each zone.
These perturbations around the equilibrium seed the parametric
instability; we would otherwise need to wait for the instability to
grow from machine roundoff error.  The vertical shearing motion $v_x =
S z$ is the local manifestation of a large scale warping motion in a
Keplerian disc, and we shall loosely refer to it as ``the warp''.

The boundary conditions are as follows.  We integrate the basic
equations (\ref{LOCMODEOM}) in a rectangular box of length $L_x$ in
the radial direction, $L_y$ in the azimuthal direction, and $L_z$ in
the vertical direction.  The grid is centred on $x = y = z = 0$.  The
radial boundaries (at $x = \pm L_x/2$) are subject to the ``shearing
box'' boundary conditions (e.g., Hawley, Gammie, \& Balbus 1995).
These require that
\begin{equation}
f(-L_x/2,y,z) = f(L_x/2,y - q\Omega L_x t,z)
\end{equation}
for $f = v_x, \delta v_y, v_z, \rho$, where $\delta v_y \equiv v_y +
q\Omega x$.  The azimuthal boundaries ($y = \pm L_y/2$) are periodic.
The vertical boundaries are reflecting.

The numerical model is integrated using a version of the ZEUS
algorithm (Stone \& Norman 1992).  ZEUS is an explicit,
finite-difference, operator-split method.  The variables lie on a
staggered mesh, so that scalar quantities are zone-centred, while
vector quantities are centred on zone faces.  It conserves mass and
linear momentum to machine roundoff error.

The ZEUS algorithm has been extensively tested (Stone \& Norman 1992;
see, e.g., Stone \& Hawley 1997 for a discussion of other
applications).  Our implementation has also been tested on a number of
standard problems.  It can, for example, reproduce standard linear
results such as sound wave propagation, and standard non-linear results
such as the Sod shock tube.

One relevant test of our shearing box implementation is uniform
epicyclic motion.  This can be initiated by setting $v_x = {\rm
  const.}$ in the initial conditions.  The fluid ought then to execute
epicyclic oscillations with period $2\pi/(\Omega (4 - 2 q)^{1/2})$.
This test is non-trivial: for a naive implementation of the Coriolis
and tidal forces the amplitude of the epicycle will grow or decay.
This is because the time-step determined from the Courant condition
depends on epicyclic phase.  Growth or decay of the oscillation can be
prevented by using ``potential velocities'' $v_{p,x} \equiv
\sqrt{v_x^2 + 4 v_y^2}$ and $v_{p,y} \equiv \sqrt{v_y^2 + v_x^2/4}$
rather than $v_x,v_y$ in the time-step condition.

A second, relevant test is the evolution of a linear amplitude bending
wave.  This test was performed in axisymmetry.  We set $L_x,L_z =
(10,10) H$, and the numerical resolution $n_x,n_z = 64,64$.  We
introduced a bending wave in the initial conditions with wavelength
$L_x$.  Linear theory gives $\omega = 0.7340\Omega$ (notice that this
is significantly different from $\Omega$, so pressure gradients play
an important role in the mode dynamics ; the mode is not a mere
sloshing up and down of the fluid in a fixed potential).  The measured
mode frequency (interval between successive zero crossings of the
vertical velocity at a fixed point in the fluid) is $\omega =
0.7336\Omega$, which differs from the true value by $1$ part in $2
\times 10^{3}$, indicating satisfactory performance on the test.

\subsection{Non-linear outcome}

The numerical model has seven important parameters:
$L_x,L_y,L_z,n_x,n_y,n_z,$ and $S$.  Before studying the effect of
these parameters we will consider a single, fiducial run in detail.

The fiducial run has an initial amplitude $S = \Omega$.  The physical
size of the box is $L_x,L_y,L_z = (4, 16, 6) H$.  The reflecting
vertical boundaries are therefore 3 scale heights away from the
mid-plane.  The numerical resolution is $n_x,n_y,n_z = 128,128,192$,
so the resolution is $32$ zones/scale-height.  The run ends at $t =
100\Omega^{-1}$.

As diagnostics, we record the evolution of three integral quantities.
The ``epicyclic energy'' is
\begin{equation}
E_{\rm epi} \equiv {1\over{2}} \int {\rm d}^3 \vx \, \rho (v_x^2 + 4 \delta
v_y^2).
\end{equation}
$E_{\rm epi}$ would be constant for the infinite wavelength warping
mode in the absence of an instability.  The vertical kinetic energy is
\begin{equation}
E_{{\rm k},z} \equiv {1\over{2}} \int {\rm d}^3 \vx \, \rho v_z^2.
\end{equation}
Absent initial perturbations, $E_{{\rm k},z} = 0$.  We expect it to grow at
twice the parasitic mode amplitude growth rate.  As a measure of the
growth of non-axisymmetric structure, we define
\begin{equation}
\NA_m \equiv \int {\rm d}x\, (c_m^2(x) + s_m^2(x)),
\end{equation}
where
\begin{equation}
c_m(x) \equiv \int {\rm d}y\, {\rm d}z\, \cos(2\pi m y/L_y) \rho v_x(\boldx).
\end{equation}
and likewise for $s_m(x)$.  $\NA_m$ will grow only if there is a
non-axisymmetric counterpart to the parametric instability or a
tertiary non-axisymmetric instability that preys on the parasitic
modes in the non-linear regime.  There are no known local, linear,
non-axisymmetric instabilities of a Keplerian disc.

Figure 1 shows the perturbed momentum field $\rho \delta\vv$ at $t =
20 \Omega^{-1}$ on a constant-$y$ slice.  Here $\delta\vv$ is the
velocity field with the parent bending wave removed.  We expect the
velocities to be dominated by the resonant inertial modes which lie
(approximately) at $60\degr$ to the vertical, and this is what is seen
in the figure.

\begin{figure}
\centerline{\epsfig{figure=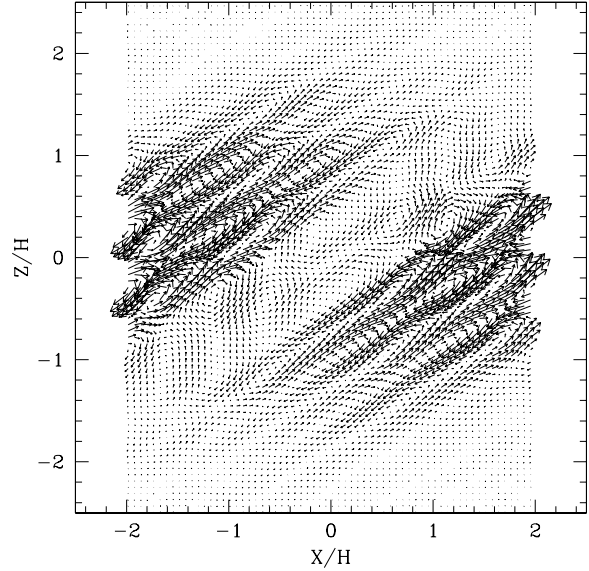, width=8.1cm}}
\caption{
The momentum field $\rho \delta\vv$ on a constant-$y$ slice
through the fiducial numerical model.  The velocity due to the parent 
bending wave has been removed.  
}
\end{figure}

The evolution of $E_{\rm epi}, E_{{\rm k},z},$ and $\NA_{1,2}$ are shown in
Figures 2, 3, and 4.  From Figure 2 it is evident that the instability
grows rapidly, removing energy from the warp.  The instability couples
strongly to compressive modes, since its characteristic velocity
amplitude is $S H = c_{\rm s} (S/\Omega) = c_{\rm s}$.  As a result
shocks are produced through which part of the energy is dissipated.
The remainder of the dissipation is via numerical averaging at the
grid scale.  The epicyclic energy is reduced by a factor of $5.5$ over
the course of the run.

\begin{figure}
\centerline{\epsfig{figure=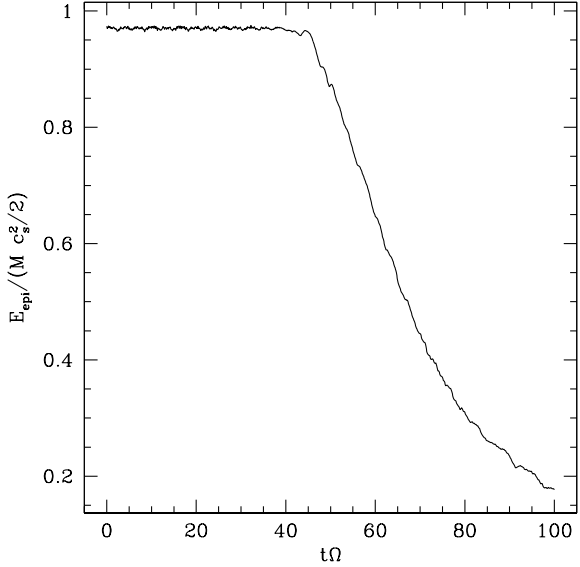, width=8.1cm}}
\caption{
Evolution of the epicyclic energy in the fiducial numerical model.
}
\end{figure}

\begin{figure}
\centerline{\epsfig{figure=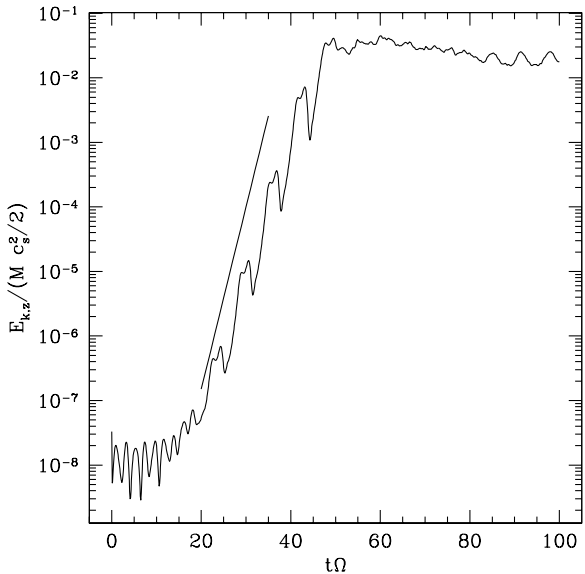, width=8.1cm}}
\caption{
Evolution of the vertical kinetic energy in the fiducial numerical
model.
}
\end{figure}

\begin{figure}
\centerline{\epsfig{figure=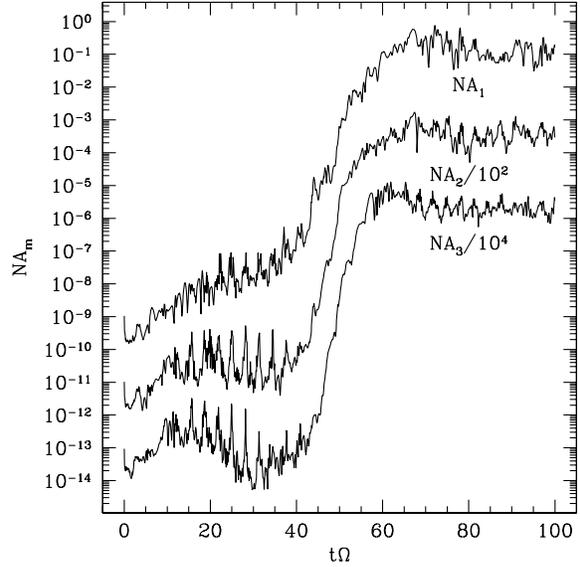, width=8.1cm}}
\caption{
Evolution, in the fiducial model, of $\NA_{1,2,3}$, which measure
lowest-order non-axisymmetric structure (see text for definition).
The ordinate units are arbitrary.
}
\end{figure}

At early times in Figure 2 a small oscillation in epicyclic energy is
evident.  This is caused by the limited accuracy of the time
integration of epicyclic motions.  As a result epicyclic energy varies
with epicyclic phase.  The oscillation is stable, however: the
epicyclic energy neither grows nor decays from cycle to cycle.  We
have confirmed that the amplitude of this oscillation decreases as
zone size and time-step decrease.

Figure 3 shows the development of $E_{{\rm k},z}$ on a logarithmic scale,
along with a line showing the analytic prediction for the growth rate.
In the linear regime, the upper envelope of $E_{{\rm k},z} \sim \exp(0.51
\Omega t)$, while the analytic theory predicts a growth rate of
$3\sqrt{3}\Omega /8 \approx 0.65 \Omega$ for the energy.  Several
factors contribute to lower the growth rate below the analytic value.
We shall discuss these below.

Figure 4 shows the development of lowest order ($m = 1,2,3$)
non-axisymmetric structure in the numerical model.  The growth rate of
$\NA_1$ is larger ($\approx 0.86\Omega$) than the growth rate of the
axisymmetric instability.  This, together with the fact that $\NA_1$
begins to grow only after the parametric instability reaches the
non-linear regime at $t \approx 45\Omega^{-1}$, suggests that, in this
case, growth of non-axisymmetric structure is due to a tertiary
non-axisymmetric instability, rather than the linear development of a
non-axisymmetric counterpart of the parametric instability.

Figure 5 shows a colour coded image of density on slices through the
model in the $x-y$ plane at $z = 0$ (top), the $x-z$ plane at $y = 0$
(middle) and the $y-z$ plane at $x = 0$ (bottom).  Black is highest
density, blue lowest.  The slices are taken late in the evolution, at
$t = 80 \Omega^{-1}$.  Density variations are visible because
compressive waves are strongly excited.  The extended, sharp features
are shocks.  The r.m.s. variation in surface density is
$\<\delta\Sigma^2\>^{1/2}/\<\Sigma\> = 0.04$.

\begin{figure}
\centerline{\epsfig{figure=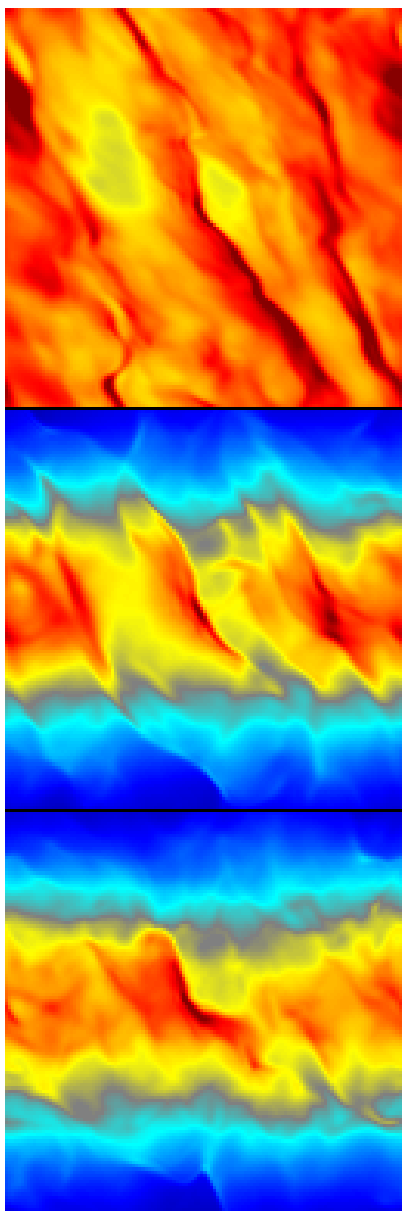, width=4.0cm}}
\vskip0.5cm
\caption{
Colour image of density on slices through the fiducial model.  The top
panel shows an $x-y$ slice at $z = 0$, the middle panel shows an $x-z$
slice at $y = 0$, and the bottom panel shows a $y-z$ slice at $x = 0$.
Black is highest density, while blue is lowest.
}
\end{figure}

\subsection{Parameter study}

We have explored the effect of each of the main model parameters on
the development of the parametric instability and its non-linear
outcome.  A full list of relevant models, with model parameters, is
given in Table 1.  Run 1 is the fiducial model.

\begin{table*}
\begin{center}
\caption{List of all numerical models}
\begin{tabular}{l|cccccccc}
No. & $L_x$ & $L_y$ & $L_z$ & $n_x$ & $n_y$ & $n_z$ & $S$ \\
\\
1 & 4 & 16  & 6 & 128 & 128 & 192 & 1 & fiducial run\\
2 & 4 & - & 6 & 256 & - & 256 & 0.5 & \\
3 & 4 & - & 8 & 340 & - & 340 & 0.5 & \\
4 & 4 & 16 & 4 & 64 & 64 & 64 & 1 & \\
5 & 4 & 16 & 10 & 64 & 64 & 160 & 1 & \\
6 & 4 & - & 10 & 64 & - & 128 & 0.5 & periodic, axisymm. \\
7 & 4 & - & 10 & 128 & - & 256 & 0.5 & periodic, axisymm. \\
8 & 4 & - & 10 & 256 & - & 512 & 0.5 & periodic, axisymm. \\
9 & 4 & - & 12 & 64 & - & 128 & 0.5 & periodic, axisymm. \\
10 & 4 & - & 12 & 128 & - & 256 & 0.5 & periodic, axisymm. \\
11 & 4 & - & 12 & 256 & - & 512 & 0.5 & periodic, axisymm. \\
12 & 4 & 8  & 6 & 64 & 32 & 96 & 1 &\\
13 & 4 & 16 & 6 & 64 & 64 & 96 & 1 \\
14 & 4 & 32  & 6 & 64 & 128 & 96 & 1 &\\
15 & 4 & 64  & 6 & 64 & 128 & 96 & 1 &\\
16 & 4 & 16  & 6 & 128 & 64 & 96 & 1 &\\
17 & 4 & 16  & 6 & 32 & 64 & 96 & 1 &\\
18 & 4 & 16  & 6 & 64 & 128 & 96 & 1 &\\
19 & 4 & 16  & 6 & 64 & 32 & 96 & 1 &\\
20 & 4 & 16  & 6 & 64 & 32 & 192 & 1 &\\
21 & 4 & 16  & 6 & 64 & 32 & 48 & 1 &\\
22 & 4 & -  & 6 & 256 & - & 256 & 1 & axisymm. \\
23 & 4 & - & 6 & 512 & - & 512 & 1 & axisymm. \\
24 & 4 & - & 6 & 256 & - & 256 & 0.05 & axisymm. \\
25 & 4 & - & 6 & 512 & - & 512 & 0.1 & axisymm. \\
26 & 4 & - & 8 & 256 & - & 256 & 0.2 & axisymm. \\
27 & 4 & - & 6 & 512 & - & 512 & 0.5 & axisymm. \\
28 & 4 & - & 6 & 256 & - & 256 & 2 & axisymm. \\
29 & 4 & - & 6 & 256 & - & 256 & 8 & axisymm. \\
\end{tabular}
\end{center}
\end{table*}

\subsubsection{Vertical box size}

One concern in formulating a numerical model such as this is whether
an artificial aspect of the boundary conditions controls the outcome.
In this respect, the azimuthal and radial boundary conditions are
somewhat less worrisome than the vertical boundary conditions; the
former restrict the scale of structure that can develop to the size of
the box, while the latter introduces a reflection of outgoing waves
that would otherwise dissipate in the disc atmosphere (although there
could be a sharp boundary between a hot disc atmosphere and the body
of the disc that would also produce reflections, albeit from a free
rather than fixed boundary).

We have tested for variation in linear growth rate with the vertical
size of box: compare, e.g., runs 2 and 3.  Both runs give ${\rm d}\ln
E_{{\rm k},z}/{\rm d} t \approx 0.25\Omega$.  Thus for values of $L_z$
close to the fiducial value, $6 H$, we observe no change in the
numerically measured growth rate.  For the non-linear development, we
observe a trend in that the decay of epicyclic energy is slower in the
larger boxes.  Quantitatively, approximately $20\%$ more of the
initial epicyclic energy is left in run 5 at $t = 75\Omega^{-1}$ than
in run 4 at the same instant.  The small size of this effect suggests
that the boundary condition is not controlling the outcome.

\subsubsection{Type of vertical boundary condition}

To test the effect of varying the character of the vertical boundary
conditions, we performed a series of axisymmetric runs with periodic
vertical boundary conditions (runs 6 to 11).  These runs are special
in that the initial conditions have
\begin{equation}
v_x = S (L_z/2\pi) \sin(2\pi z/L_z) + \delta
\end{equation}
to make them compatible with the boundary conditions.  This velocity
profile then requires $L_z$ larger than reflecting boundary condition
runs so as to better approximate a linear shear near the mid-plane of
the disc; this in turn implies lower numerical resolution $n_z/L_z$
for a given grid size $n_z$.

The growth rates measured in the periodic vertical runs are not
distinguishably different from those measured in reflecting boundary
condition runs at similar resolution.  The decay of epicyclic energy
also follows a similar trajectory under the two different boundary
conditions.  This again suggests that the boundary conditions are not
controlling the outcome.

\subsubsection{Azimuthal box size}

One might be concerned that the limited azimuthal size of the box is
somehow limiting the linear development of the instability or the
non-linear outcome.  This concern turns out to be justified over a
limited range in $L_y$ for both the linear and non-linear development
of the instability.

Figure 4 shows the evolution of $\NA_{1,2,3}$ in the fiducial run 1.
This demonstrates that $m = 1$ structure grows most quickly, and
suggests that the low-$m$ modes available in a larger box might grow
even more quickly.  Figure 6 demonstrates that this is indeed the
case.  It shows the evolution of $\NA_1/L_y$ in runs 12, 13, 14, and
15, which have $L_y = 8,16,32,64$ respectively.  The numerical
resolution $n_y/L_y$ is constant between these runs.  The early growth
of $\NA_1$ is strongest for $L_y = 64$.  In fact, the growth rate of
$\NA_1$ for $L_y = 64$ is $0.32\Omega$, close to the $0.42\Omega$
measured for ${\rm d}\ln E_{{\rm k},z}/{\rm d}t$.  This evidence
suggests that there is a non-axisymmetric counterpart of the
parametric instability for, and only for, sufficiently low $m/L_y$.

\begin{figure}
\centerline{\epsfig{figure=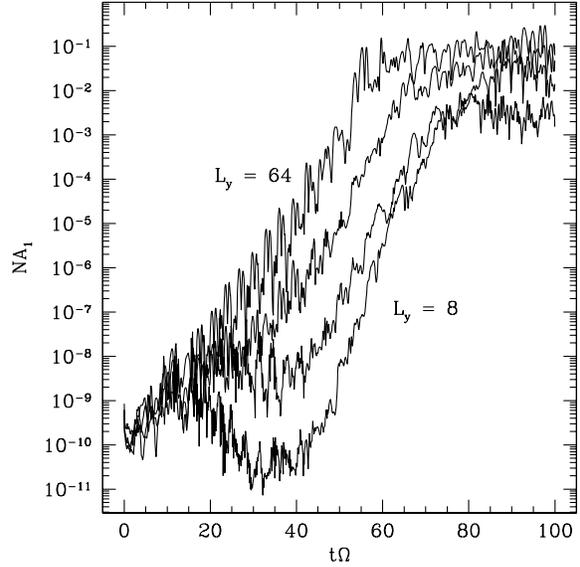, width=8.1cm}}
\caption{
Evolution of $\NA_1/L_y$ in runs with $L_y = 8, 16,32,$ and $64$.
The ordinate units are arbitrary.
}
\end{figure}

In retrospect this result is easy to understand.  Our linear analysis
could be trivially generalized to weakly non-axisymmetric disturbances
using the shearing wave formalism of non-axisymmetric density wave
theory (Goldreich \& Lynden-Bell 1965).  The radial wavenumber of the
disturbance would then evolve according to $\dot{k}_x = q\Omega k_y$.
The amplitudes of these shearing waves will grow so long as the radial
wavenumber does not change so much over a growth time that the
resonance between the daughter modes and parent mode is detuned.

The non-linear development of the instability is not strongly altered
by variation of $L_y$.  There is a tendency for $E_{\rm epi}$ to decay
slightly more slowly in models with larger $L_y$.  Presumably this is
because large-scale, slowly dissipating motions are available in the
large $L_y$ box, and these motions act as a reservoir for the
epicyclic energy.

\subsubsection{Numerical resolution}

Runs 16 through 21, 1, 22, and 23 test the effect of numerical
resolution.  The measured linear growth rate is slightly dependent on
resolution in the meridional plane.  For $n_x,n_z = 32,48$,
$\gamma_{\rm num} \equiv {\rm d}\ln E_{{\rm k},z}/{\rm d}t =
0.37\Omega$; for $n_x,n_z = 64,96$, $\gamma_{\rm num} = 0.46\Omega$;
for $n_x,n_z = 128,192$, $\gamma_{\rm num} = 0.49\Omega$; for $n_x,n_z
= 512,512$, $\gamma_{\rm num} = 0.53\Omega$.  This should be compared
to the linear theory rate $\gamma = 0.65\Omega$.  The non-linear decay
rate of the warp also has a trend with resolution in that high
resolution models decay more slowly.

\subsubsection{Initial warp amplitude}

At small $S/\Omega$, the three-mode coupling analysis of \S 5 predicts
that
\begin{equation}
{{\rm d}\ln E_{{\rm k},z}\over{{\rm d} t}}  =
{3\sqrt{3} S\over{8}} \approx 0.65 S.
\end{equation}
To test this, we measured a numerical growth rate for a series of
values of $S$.  In each case, we used the run with the highest
available numerical resolution in the meridional plane (we will be
concerned only with the development of axisymmetric instabilities
here).  The results are listed in Table 2.  The numerical growth rates
typically lie $\approx 15\%$ below the analytic rates.  We believe
that this is primarily a resolution effect, since the growth rate
grows monotonically with increasing resolution.

At large $S/\Omega$ the three-mode coupling analysis is no longer
applicable.  The incompressible, unstratified model of \S 4 is not
limited to small $S/\Omega$, and numerical integrations of those model
equations show reduced growth for $S/\Omega \gta 1$.  This is observed
in the numerical models listed in Table 2.

\begin{table}
\begin{center}
\caption{List of measured and predicted values for growth rate}
\begin{tabular}{l|cccc}
No. & $S/\Omega$ & $\gamma_{\rm num}/\Omega$ & $\gamma_{\rm th}/\Omega = 
3\sqrt{3} S/8$ & $\gamma_{\rm num}/\gamma_{\rm th}$ \\
\\
24 & 0.05 & 0.020 & 0.032 & 0.62 \\
26 & 0.1  & 0.055 & 0.065 & 0.85 \\
26 & 0.2  & 0.11  & 0.13  & 0.85 \\
27 & 0.5  & 0.28  & 0.32  & 0.87 \\
23 & 1.0  & 0.49  & 0.65  & 0.76 \\
28 & 2.0  & 0.82  & 1.3   & 0.63 \\
29 & 8.0  & 1.7   & 5.2   & 0.33 \\
\end{tabular}
\end{center}
\end{table}

\section{Discussion}

We have shown that warps in unmagnetized, non-self-gravitating,
inviscid Keplerian discs are subject to an instability that leads, in
the non-linear regime, to damping of the warp.  The damping is
non-linear in that the growth rate of the instability is proportional
to the amplitude of the warp.  How might this process influence the
development of warps in astrophysical discs?

A full answer to this question awaits an understanding of other
damping mechanisms that might act on a warp.  If the disc is already
turbulent, the parametric instability may not occur.  In the Appendix,
we estimate the influence of an isotropic effective turbulent
viscosity on our analysis.  We find that the instability still occurs
if the amplitude of the in-plane oscillations is sufficiently large,
possibly $S/\Omega\ga30\alpha$, where $\alpha$ is the dimensionless
viscosity parameter.  In addition, for a given warping of the disc,
$S/\Omega$ is reduced in the presence of viscosity because the
resonance that drives the in-plane oscillations is weakened.
Therefore our analysis is most relevant to discs with $\alpha\ll1$.

Assuming that the instability develops into the non-linear regime, we
may hazard a guess at the consequences for the warp, using a simple
model for the interaction of in-plane and out-of-plane motions in the
warp.  We then apply the result to the warped, masing disc in NGC
4258.

Consider a linear, axisymmetric bending wave in the corotating frame.
The radial and vertical displacements associated with the wave, $\xi_r
\sim z \exp({\rm i} k r)$ and $\xi_z \sim \exp({\rm i} k r)$, obey
\begin{equation}\label{OSCR}
{{\rm d}^2 \xi_r\over{{\rm d}t^2}} + \kappa^2 \xi_r = 
- {\rm i} k \xi_z \Omega^2 z,
\end{equation}
\begin{equation}\label{OSCZ}
{{\rm d}^2 \xi_z\over{{\rm d}t^2}} + \Omega^2 \xi_z = 
{\rm i} k c_{\rm s}^2 \xi_r \del_z \xi_r.
\end{equation}

It is evident that the in-plane and out-of-plane motions behave as
coupled harmonic oscillators.  If $k \sim r^{-1}$ and $H/r \ll 1$, as
we shall assume, then the coupling is weak in the sense that $k c_{\rm s}
\ll \Omega$.  The coupling would be negligible except that $\kappa^2
\simeq \Omega^2$ for a Keplerian disc, so the vertical and radial
oscillations are resonant.

We will use equations (\ref{OSCR},\ref{OSCZ}) as the basis of a simple
model for the warp.  In addition to the terms from linear theory, we
will include an out-of-plane driving term, meant to model the Pringle
instability, and a non-linear damping term, meant to model the effects
of the parametric instability.  In terms of the normalized
displacements $X_z = \xi_z/H$ and $X_r = {\rm i} \xi_r(z = H)/H$, the
governing equations are
\begin{equation}\label{OSCZ1}
{{\rm d}^2 X_z\over{{\rm d}t^2}} + \Omega^2 X_z =
- \eps\, \Omega^2 X_r + \gamma \Omega {{\rm d} X_z\over{{\rm d} t}},
\end{equation}
\begin{equation}\label{OSCR1}
{{\rm d}^2 X_r\over{{\rm d}t^2}} + \kappa^2 X_r = - \eps\, \Omega^2 X_z 
	- f(\dot{X}_r).
\end{equation}
Here $\eps \equiv k c_{\rm s}/\Omega$, $\gamma/2$ is a dimensionless growth
rate for the driving instability, and $f$ is some function of
$\dot{X}_r\equiv {\rm d}X_r/{\rm d}t$ that models the damping effect of the
parametric instability on in-plane motions.

We set $\kappa=\Omega=1$ and look for steady-state oscillatory
solutions of equations (\ref{OSCZ1}, \ref{OSCR1}).  To start, define
``action-angle'' variables by
\begin{eqnarray}
X_z &\equiv& a_z\cos\theta_z, ~~~~\dot X_z \equiv -a_z\sin\theta_z,\\
X_r &\equiv& a_r\cos\theta_r,~~~~\dot X_r \equiv -a_r\sin\theta_r.
\end{eqnarray}
Consistency of these definitions requires
\begin{eqnarray}
\dot a_z\cos\theta_z-\dot\theta_z a_z\sin\theta_z &=& -a_z\sin\theta_z,\\
\dot a_r\cos\theta_r-\dot\theta_r a_r\sin\theta_r &=& -a_r\sin\theta_r.
\end{eqnarray}
These relations and the differential equations for $X_r$ and $X_z$
yield
\begin{eqnarray}
\dot a_z &=&\gamma a_z\sin^2\theta_z +\eps a_r\cos\theta_r\sin\theta_z,\\
\dot a_r &=& f \sin\theta_r +\eps a_z\sin\theta_r\cos\theta_z,\\
\dot\theta_z &=& 1 +\gamma\sin\theta_z\cos\theta_z 
+ {\epsilon a_r\over a_z} \cos\theta_r\cos\theta_z,\\
\dot\theta_r &=& 1 + \frac{f}{a_r} \cos\theta_r 
+ {\epsilon a_z\over a_r}\cos\theta_r\cos\theta_z.
\end{eqnarray}
When $\gamma$, $\eps$, and $f$ are $\ll 1$, the rapidly-oscillating
terms can be averaged out:
\begin{eqnarray}
\Delta\dot \theta &\approx& {\eps\over{2}}\left({a_z\over a_r}-{a_r\over a_z}
\right)\cos\Delta\theta,\\
\dot a_z &\approx& {\gamma\over2}a_z
-{\epsilon\over2}a_r\sin\Delta\theta,\\
\dot a_r &\approx& \bar{f}(a_r)
+{\epsilon\over2}a_z\sin\Delta\theta,
\end{eqnarray}
where $\Delta\theta\equiv\theta_r-\theta_z$, $\bar{f} \equiv \<
f\sin\theta_r\>$, and $\<\>$ denotes a phase average.  Since $f$
should be an odd function of $\dot{X}_r$, we have $\langle
f\cos\theta_r\rangle=0$.

A steady state requires either $a_z = a_r$ or $\cos\Delta\theta = 0$.
In astrophysical applications the latter solutions are likely to be
more relevant, because thin discs with observable warps have $a_z \gg
1$, while $a_r \gg 1$ would require implausibly supersonic motions
within the disc.

The steady state solutions with $\cos\Delta\theta = 0$ have
\begin{eqnarray}
{\bar{f}(a_r)\over{a_r}} & = & -{\eps^2\over{2\gamma}},\\
a_z & = & (\eps/\gamma) a_r.\label{az}
\end{eqnarray}
Stability of this solution requires
\begin{equation}
a_r < a_z 
\end{equation}
and, for $n \equiv {\rm d}\ln\bar{f}/{\rm d}\ln a_r$,
\begin{equation}
{\gamma^2\over{\eps^2}} < n < 1.
\end{equation}
To go further, we must guess at $\bar{f}(a_r)$.  At very small $a_r$,
we expect $n \ge 2$ because the growth rate of the parametric
instability is proportional to $a_r$.  At larger amplitudes $f$ models
the effect of stationary, fully developed turbulence.  If turbulence
initiated by the parametric instability acts as a viscosity (as might
also be expected in an MHD turbulent disc) then $n \approx 1$.  Notice
that $n \approx 1$ is consistent with the approximately exponential
damping of the warp observed in the numerical experiments of \S 6.
The decay time of the epicyclic energy shown in Figure 2 is $\approx
30\Omega^{-1}$, which could be fitted by $f\approx a_r/60$: i.e., by a
dimensionless viscosity $\alpha\approx 1.7\times 10^{-2}$.

Let us apply these results to the masing disc in NGC 4258 (Miyoshi et
al. 1995, also Gammie et al. 1999 and references therein).  The
presence of water vapour suggests $c_{\rm s} \sim 1 \kms$.  The rotational
velocity is $v_c \approx 1100 (r/ 0.13\pc)^{-1/2} \kms$.  Thus $\eps
\sim H/r \approx 10^{-3}$.  The orientation of the inner edge of the
masing disc at $\approx 0.13\pc$ differs from that at the outer edge
at $\approx 0.24 \pc$ by $\approx 0.25$ radians, so $a_z \approx 250$.
The corresponding velocity of in-plane motions at $z = H$ would be
$\approx 250\kms$, were linear theory applicable.  Finally, an
estimate of the growth time of the Pringle instability gives $\tau
\sim 5 \times 10^6 \yr$ \cite{mbp96}, or $\gamma\equiv 2/(\Omega\tau)
\sim 1.2 \times 10^{-4}$ if $\Omega$ is taken at the outer edge of the
disc.  Direct measurements of maser velocities show that the departure
from a Keplerian rotation curve is small, so we may take $\kappa =
\Omega$.

Of the model parameters $\eps, a_z,$ and $\gamma$, the last is most
uncertain.  Let us assume that the disc is described by our model
equations and is in equilibrium, then solve for $\gamma$.  Suppose $f
\approx \alpha\, {\rm sign}(\dot{X_r}) |\dot{X_r}|^n/X_0^{n-1}$, where
$X_0$ is a characteristic amplitude, and $n$ is just slightly smaller
than $1$ so the equilibrium is stable.  Here $\alpha$ is the
dimensionless strength of the damping.  For $n = 1$ the damping rate
is precisely what one would expect for a viscous disc with $\nu =
\alpha c_{\rm s}^2/ \Omega$.  Then $\bar{f} = - \alpha p_{n+1} {\rm
  sign}(a_r) |a_r|^n/X_0^{n-1}$, where $p_{n+1} \equiv \langle
|\sin\theta_r|^{n+1} \rangle$, and
\begin{equation}
\gamma = (2 \alpha p_{n+1})^{-1/n} \eps^{1 + 1/n} |a_z|^{-1 + 1/n}
        X_0^{1 - 1/n}.
\end{equation}
In the limit $n \rightarrow 1$, $\gamma \rightarrow \eps^2/\alpha =
10^{-6}/\alpha$.  This is smaller than the nominal value by a factor
of $100 \alpha$.

We can obtain an upper limit on $\gamma$ if we suppose that the warp
is limited by coupling to in-plane motions and that $a_r < 1$.  Then,
to satisfy equation (\ref{az}), we must have $\gamma < \eps/a_z \approx 4
\times 10^{-6}$.  This limit is due to the weakness of coupling
between in-plane and out-of-plane motions.  For larger growth rates
the warp must be limited by some other process, or perhaps it is not
limited and the disc is destroyed.  The maximum dissipation rate to be
expected of \emph{any} local mechanism (one that relies upon the
internal dynamics of the disc) is $\sim c_{\rm s}^2\Omega$ per unit mass.
The local energy in vertical motions of the warp as measured in the
corotating frame of the gas is $E_{\rm w}\sim \Omega^2\langle
\xi_z^2\rangle$ per unit mass, and the rate at which the Pringle
instability adds to this energy is
\begin{displaymath}
\gamma\Omega E_{\rm w} \sim 6\left(\frac{\gamma}{10^{-4}}\right)
\left\langle\left(\frac{\xi_z}{250 H}\right)^2\right\rangle~c_{\rm s}^2\Omega.
\end{displaymath}
It is therefore difficult to see how any local mechanism can limit the
warp if the growth rate is as large as Maloney et al.'s estimate.

Part of this discrepancy might be explained by uncertainties in the
parameters of Maloney et al.'s model.  It is also possible that
non-linear effects reduce the growth rate of the Pringle instability.
There is some evidence for this in that the disc in NGC 4258 is not as
strongly twisted as one would expect from linear theory (see
Herrnstein, 1997 for a warp model that fits the observations).
Finally, non-linear effects may also enhance the coupling $\eps$
between in-plane and out-of-plane oscillations.

JG and CFG are supported by NASA Origins grant NAG5-8385.  GIO is
supported by Clare College, Cambridge.  We thank the Isaac Newton
Institute for their support during the programme on Dynamics of
Astrophysical Discs, where this work was initiated.  We also thank
John Papaloizou and Caroline Terquem for their comments.

\appendix

\section{Effect of viscosity on the instability}

Here we make a rough estimate of the influence of pre-existing
turbulence on the development of the parametric instability.  In the
absence of a proper understanding of the interaction of the in-plane
oscillations with turbulence (see Torkelsson et al. 2000 for a direct
numerical approach) we suppose that the process may be treated as an
isotropic effective viscosity of the form
\begin{equation}
  \nu=\alpha c_{\rm s}^2/\Omega=\alpha \Omega H^2,
\end{equation}
where $\alpha$ is the usual dimensionless viscosity parameter.

The Floquet analysis of \S 4 is easily adapted to include viscosity.
The solution is identical but multiplied by the damping factor
\begin{equation}
  \exp\left(-\int\nu k^2\,{\rm d}t\right).
\end{equation}
When $S/\Omega$ is small the time-dependence of $\boldk$ may be
neglected and the growth rate of the instability is simply reduced
according to
\begin{equation}
  s\mapsto s-\nu k^2.
\end{equation}
For a tuned mode with $k_x/k_z=\pm\sqrt{3}$, the modified growth
rate is
\begin{equation}
  s\approx{{3\sqrt{3}}\over{16}}S-4(k_z H)^2\alpha\Omega.
\end{equation}
In the unstratified model $k_z$ is unrestricted, but, in reality, the
mode must fit inside the disc.  An approximate criterion can be
obtained by requiring that at least half a wavelength fit into a
vertical distance $2H$, i.e. $k_z H\ga\pi/2$.  The instability then
develops provided that
\begin{equation}
  {{S}\over{\Omega}}\ga 30\alpha.
  \label{criterion}
\end{equation}

The effect of a small viscosity on the mode-coupling analysis of \S 5
can also be estimated.  Provided that bulk viscosity effects may be
neglected, the eigenfunctions of wave modes in an isothermal disc (\S
3) can still be obtained in terms of Hermite polynomials when
viscosity is included.  The dispersion relation is much more
complicated, but can be expanded for small $\alpha$ to obtain
\begin{equation}
  \omega_n(k)=\omega_n^{(0)}(k)-{\rm i}\alpha\Omega d_n(k)+O(\alpha^2),
\end{equation}
where the first term is the frequency in the inviscid case (equation
\ref{dispersion}) and $d_n$ is a dimensionless measure of the viscous
damping rate of the mode.  Generally speaking, $d_n$ increases with
increasing $n$ and $k$.  We therefore consider the first parametric
resonance, which occurs between modes 1 and 2 (i.e.
$\omega_1+\omega_2=\Omega$) at $kH\approx 1.88$.  The inviscid growth
rate is then $s\approx 0.138S$.  At this point $d_1\approx d_2\approx
4.5$, giving a net growth rate of $0.138S-4.5\alpha\Omega$.  The
instability proceeds only if $S/\Omega\ga33\alpha$, which agrees
closely with equation (\ref{criterion}).


\begin{thebibliography}{}

\bibitem[\protect\citename{Abramowitz \& Stegun }1965]{as65}
	Abramowitz M., Stegun I. A., 1965, Handbook of Mathematical
	Functions, Dover, New York
\bibitem[\protect\citename{Balbus \& Hawley }1998]{bh98}
	Balbus S. A., Hawley J. F., 1998, Rev. Mod. Phys., 70, 1
\bibitem[\protect\citename{Bayly }1986]{b86}
	Bayly B. J., 1986, Phys. Rev. Lett., 57, 2160
\bibitem[\protect\citename{Bender \& Orszag }1978]{bo78}
	Bender C. M., Orszag, S. A. 1978, Advanced Mathematical
	Methods for Scientists and Engineers, McGraw-Hill, New York
\bibitem[\protect\citename{Craik \& Criminale }1986]{cc86}
	Craik A. D. D., Criminale W. O., 1986,
	Proc. R. Soc. Lond. A, 406, 13
\bibitem[\protect\citename{Gammie }1996]{g96}
	Gammie C. F., 1996, ApJ, 457, 355
\bibitem[\protect\citename{Gammie \& Menou }1998]{gm98}
	Gammie C. F., Menou K., 1998, ApJ, 492, 75
\bibitem[\protect\citename{Gammie et al.}1999]{gnb99}
	Gammie C. F., Narayan, R., \& Blandford, R. 1999, 
	ApJ, 516, 177
\bibitem[\protect\citename{Goldreich \& Lynden-Bell }1965]{glb65}
	Goldreich P., Lynden-Bell D., 1965, MNRAS, 130, 125
\bibitem[\protect\citename{Goodman }1993]{g93}
	Goodman J., 1993, ApJ, 406, 596
\bibitem[\protect\citename{Hawley, Gammie \& Balbus }1995]{hgb95}
	Hawley J. F., Gammie C. F., Balbus S. A., 1995, ApJ, 440, 742
\bibitem[\protect\citename{Herrnstein }1997]{herr97}
	Herrnstein, J. 1997, Ph.D. Thesis, Harvard University
\bibitem[\protect\citename{Katz }1973]{k73}
	Katz J. I., 1973, Nat, 246, 87
\bibitem[\protect\citename{Kelvin }1887]{k87}
	Lord Kelvin, 1887, Phil. Mag., 24, 188
\bibitem[\protect\citename{Kumar \& Goodman }1996]{kg96}
	Kumar P., Goodman J., 1996, ApJ, 466, 946
\bibitem[\protect\citename{Kumar \& Coleman }1993]{kc93}
	Kumar S., Coleman C. S., 1993, MNRAS, 260, 323
\bibitem[\protect\citename{Lubow \& Pringle }1993]{lp93}
	Lubow S. H., Pringle J. E., 1993, ApJ, 409, 360
\bibitem[\protect\citename{Maloney et al. }1996]{mbp96}
	Maloney, P., Begelman, M., \& Pringle, J. 1996, ApJ, 472, 582
\bibitem[\protect\citename{Maoz }1995]{m95}
	Maoz, E. 1995, ApJ, 447, L91
\bibitem[\protect\citename{Miyoshi et~al. }1995]{mmhgndi95}
	Miyoshi M., Moran J., Herrnstein J., Greenhill L., Nakai N.,
	Diamond P., Inoue M., 1995, Nat, 373, 127
\bibitem[\protect\citename{Ogilvie }1999]{o99}
	Ogilvie G. I., 1999, MNRAS, 304, 557
\bibitem[\protect\citename{Ogilvie \& Lubow }1999]{ol99}
	Ogilvie G. I., Lubow S. H., 1999, ApJ, 515, 767
\bibitem[\protect\citename{Papaloizou \& Lin }1995]{pl95}
	Papaloizou J. C. B., Lin D. N. C., 1995, ApJ, 438, 841
\bibitem[\protect\citename{Papaloizou \& Terquem }1995]{pt95}
	Papaloizou J. C. B., Terquem C., 1995, MNRAS, 274, 987
\bibitem[\protect\citename{Pringle }1996]{p96}
	Pringle J. E., 1996, MNRAS, 281, 357
\bibitem[\protect\citename{Roberts }1974]{r74}
	Roberts W. J., 1974, ApJ, 187, 575
\bibitem[\protect\citename{Sagdeev \& Galeev }1969]{sg69}
	Sagdeev R. Z., Galeev A. A., 1969, Nonlinear plasma theory, 
	Benjamin, New York
\bibitem[\protect\citename{Stone \& Norman }1992]{sn92}
	Stone J. M., Norman M. L., 1992, ApJS, 80, 753
\bibitem[\protect\citename{Tananbaum et~al. }1972]{tgklsg72}
	Tananbaum H., Gursky H., Kellogg E. M., Levinson R.,
	Schreier E., Giacconi R., 1972, ApJ, 174, L143
\bibitem[\protect\citename{Torkelsson et~al. }2000]{tobpns00}
	Torkelsson U., Ogilvie G. I., Brandenburg A., Pringle J. E.,
	Nordlund \AA, Stein R. F., 2000, MNRAS, in press

\end{thebibliography}
\end{document}